\documentstyle[aps,pre,twocolumn]{revtex}
\begin{document}
\title{Hamiltonian for a restricted isoenergetic thermostat}
\author{C. P. Dettmann}
\address{Rockefeller University, 1230 York Ave, New York NY 10021}
\date{\today}
\maketitle
\begin{abstract}
Nonequilibrium molecular dynamics simulations often use mechanisms called
thermostats to regulate the temperature.  A Hamiltonian is presented for
the case of the isoenergetic (constant internal energy) thermostat
corresponding to a tunable isokinetic (constant kinetic energy) thermostat,
for which a Hamiltonian has recently been given.
\end{abstract}

Thermostats are modifications to the equations of motion of a classical
system to simulate thermal interaction of a system with the environment.
The Nos\'e-Hoover thermostat is used to simulate fluctuations in energy
of an equilibrium system corresponding to the canonical ensemble of
statistical mechanics, and the Nos\'e-Hoover and Gaussian thermostats,
among others, are used to remove heat from a
system driven by external forces into a nonequilibrium stationary
state~\cite{MD}.  There has been a recent interest in thermostatted
equations of motion, focussed on the symplectic structure of the equations
of motion, and the related pairing of the Lyapunov exponents.  Both a
Hamiltonian and pairing of Lyapunov exponents are known for Nos\'e-Hoover
and Gaussian isokinetic (GIK: constant kinetic energy)
thermostats~\cite{MD,WL,DM}.
Numerical evidence against pairing (and hence the existence of a Hamiltonian)
are discussed in~\cite{SEI} for the GIK thermostat
applied to shearing systems and in~\cite{BCP} for the Gaussian
isoenergetic (GIE: constant internal energy) thermostat.  The latter
paper does, however show that in a special case of the GIE thermostat,
involving one rather than two arbitrary potentials,
the Lyapunov exponents are paired.  The purpose
of this short note is present a Hamiltonian for this case.

The GIE thermostat has equations of motion of the form
\begin{eqnarray}
\frac{d{\bf x}_i}{dt}=\frac{{\bf p}_i}{m_i}\label{e:x}\;\;,& &\quad
\frac{d{\bf p}_i}{dt}=-\frac{\partial\Phi^{(ext)}}{\partial {\bf x}_i}
-\frac{\partial\Phi^{(int)}}{\partial {\bf x}_i}-\alpha{\bf p}_i
\;\;,\nonumber\\
\alpha&=&-\frac{\sum_i \frac{{\bf p}_i}{m_i}\cdot
\frac{\partial\Phi^{(ext)}}{\partial {\bf x}_i}}
{\sum_i {\bf p}_i\cdot{\bf p}_i/m_i}\;\;,\label{e:alpha}
\end{eqnarray}
where $\Phi^{(ext)}$ is the external driving potential, $\Phi^{(int)}$ the
interparticle potentials, and $\alpha$ is the thermostat term which ensures
that the equations conserve internal energy
$E=\sum_i{\bf p}_i^2/(2m_i)+\Phi^{(int)}$.  The equations reduce to no
thermostat when $\Phi^{(ext)}=0$ and to GIK when $\Phi^{(int)}=0$.  A more
general example of a limit involving only one arbitrary potential is the
case $\Phi^{(ext)}=\gamma\Phi$, $\Phi^{(int)}=(1-\gamma)\Phi$, leading
to the equations
\begin{equation}\label{e:F}
\frac{d{\bf x}_i}{dt}=\frac{{\bf p}_i}{m_i}\;\;,\quad
\frac{d{\bf p}_i}{dt}=-\frac{\partial\Phi}{\partial {\bf x}_i}
+\gamma\frac{\sum_i \frac{{\bf p}_i}{m_i}\cdot
\frac{\partial\Phi}{\partial {\bf x}_i}}
{\sum_i {\bf p}_i\cdot{\bf p}_i/m_i}{\bf p}_i\;\;,
\end{equation}
which conserve energy $E=\sum_i{\bf p}^2_i/(2m_i)+(1-\gamma)\Phi$.
Here, $\gamma$ effectively controls the strength of the thermostat
from no thermostat ($\gamma=0$), to the GIK thermostat ($\gamma=1$).
For any $\gamma$ the Lyapunov exponents
are paired~\cite{BCP}, suggesting the existence of a Hamiltonian.

Following the GIK case~\cite{DM}, the conservation
law is enforced by setting the numerical value of the Hamiltonian equal to
the conserved energy, assigned the value zero by a
shift in the potential energy.  This allows the kinetic energy term in the
denominator of~(\ref{e:alpha}) to be replaced by minus the potential
energy (note $\Phi<0$),
\begin{equation}
\alpha=\frac{\gamma}{2(1-\gamma)}\sum_i\frac{{\bf p}_i}{m_i}\cdot
\frac{\partial}{\partial{\bf x}_i}\ln|\Phi|\;\;.
\end{equation}

Another aspect of a Hamiltonian description of thermostatted systems is that
in the physical variables $({\bf x},{\bf p})$ there is a phase space volume
contraction rate proportional to $\alpha$, while in the canonical variables
$({\bf x},\mbox{\boldmath{$\pi$}})$ phase space volume is conserved.  This
means that $\mbox{\boldmath{$\pi$}}$ must be greater than ${\bf p}$
by a factor equal to
$\exp\left(\int\alpha dt\right)=|\Phi|^{\gamma/(2(1-\gamma))}$.  Multiplying
the zero energy by an arbitrary power of $|\Phi|$ we have
\begin{equation}
H_\beta({\bf x},\mbox{\boldmath{$\pi$}},\lambda)=
|\Phi|^{-\frac{\gamma}{1-\gamma}+\beta}
\sum_i\frac{\mbox{\boldmath{$\pi$}}_i^2}{2m_i}
+(1-\gamma)\Phi|\Phi|^{\beta}\;\;,
\end{equation}
which, combined with the constraint $H_\beta=0$ and the identifications
$dt=|\Phi|^{-\gamma/(2(1-\gamma))+\beta}d\lambda$ and
${\bf p}_i=|\Phi|^{-\gamma/(2(1-\gamma))}\mbox{\boldmath{$\pi$}}_i$ leads
to the equations of motion,~(\ref{e:F}).
Interesting cases are $\beta=\gamma/(2(1-\gamma))$ for
which there is no time scaling, $\beta=0$ has a certain simplicity,
$\beta=-\gamma/(1-\gamma)$ yields the familiar
form of kinetic plus potential energy, and $\beta=-1$ for which the
Hamiltonian is that of a geodesic in a conformally flat space,
see~\cite{DM}.

The author is grateful for discussions with W. G. Hoover.

\end{document}